# Abnormal heavy-flux flow in cuprate superconductors

Xi Chen[1], Jianghui Pan[2], Jiahao Dong[3], Changkai Luo[2], Yuchuang Cao[1], Xin Li[1, †]

1. John A. Paulson School of Engineering and Applied Sciences, Harvard University, Cambridge, MA 02138, United States

2. School of Physics, Peking University, Beijing 100871, China

3. Department of Materials Science and Engineering, Massachusetts Institute of Technology, Cambridge, MA 02139, United States

**Abstract**

Dynamic charge transfers, or charge fluxes, generated by anharmonic phonon couplings have previously shown some fundamental connections with the pseudogap phase and other important phases in cuprate superconductors. [1,2] The anharmonicity here suggests the importance of nonadiabatic behaviors of charge fluxes in cuprates. [2,3] These charge fluxes will result in the dynamic charge transfer that can be directly calculated by the charge difference induced by phonons in density functional theory simulations. [2] However, retrieving the flux field from the calculated charge transfer is nontrivial. In this article, a new method is developed to solve these fluxes, which help identify anharmonically coupled phonons as the generator of nonlocal flux flows beyond local oscillations in cuprates, especially at high frequencies. A strength metric of such nonlocal fluxes shows a strong correlation to the maximum superconducting transition temperature $T_{c,max}$ of different cuprate families. A unique picture is thus proposed where heavy fluxes tunnel through a dynamic energy band near the Fermi level, providing the attraction force through a Bernoulli-like mechanism for the formation of Cooper pairs. Adding the dynamic band to our band simulation also reproduces the flat band feature that appears below the superconducting transition temperature $T_c$ in previous experimental measurements.

†: Corresponding author: lixin@seas.harvard.edu

**Introduction**

The pairing mechanism in cuprate superconductors is not only a paramount question in condensed matter physics, but also crucial to the design of higher $T_c$ superconductors at ambient pressure. Recently, more experiments have suggested the importance of dynamic behaviors involving the coupled ultrafast motions of ions and charges. [4,5] Our density functional theory (DFT) simulations also found that higher apical oxygen oscillation frequency and stronger apical charge flux are positively correlated with $T_{c,max}$ among a wide range of cuprate families. [1]

More recently, charge density waves and charge fluctuations in cuprate superconductors attracted increasing interests. [6–12] We further show that the charge fluxes generated by anharmonically coupled phonons may play a crucial role in understanding their pseudogap phase and other important phases. [2] However, in our recent work [2] only the result of fluxes, that is the charge transfer Δρ, was analyzed, while the direction information of most fluxes was still ambiguous, because it lacked a systematic method to analyze such a flux behavior.

In this article, we first formulate a minimization constraint to uniquely determine the flux $\vec{j}$ associated with each atomic bond from the entire Δρ distribution in a supercell induced by phonons. Using this method, we can obtain all the fluxes of different atomic bonds generated by all the anharmonically coupled phonons in DFT simulations, giving the so-called flux patterns. The similarity between the two flux patterns at the two extrema of phonon oscillations in a period thus describes different types of flux-phonon interactions. A very negative similarity suggests an immediate change of directions for local charge transfers between most neighboring atoms generated by phonons, which is a conventional picture of local flux oscillations satisfied by many materials based on our analysis, including many Fe-based and BCS superconductors. Surprisingly, we find that for cuprate superconductors with the highest $T_c$ at ambinent pressure, and $H_3S$ with even higher $T_c$ at high external pressure as well, the similarity is very positive. This suggests an abnormal flux-phonon interaction, where most local charge transfers do not need to change the direction with phonon oscillations. This forms the intrinsic foundation for nonlocal flux flow oscillations, which is a phenomenon that can be further magnified when electron or hole dopings are introduced as extrinsic source to provide much stronger charge fluxes. The finding thus unveils that anharmonic phonon pairs may play a unique role as the generator of nonadiabatic flux flows in these materials. More interestingly, we find that cuprates with higher $T_c$ show larger strength of such fluxes. Especially, the Ba or Ca ions that were considered as inactive in conventional pictures are found to be a critical hub to regulate such fluxes.

These real space fluxes also reflect nontrivial dynamic oscillations of energy bands in the reciprocal space. By introducing the dynamic energy band associated with these fluxes, a three-band Hamiltonian is constructed to describe the behavior of cuprates near and below $T_c$, where the added third band here to the previous two-band model [2] is purely dynamic in nature and is flat in shape. Fluxes in such a dynamic band are thus at least 10 times heavier than the free electron mass and around 10 times slower than the electron Fermi velocity in cuprates, with the maximum flux speed being set by the highest frequency of phonon modes. Angle-resolved photoemission spectroscopy (ARPES) simulation based on the three-band model well reproduces the flat band feature observed in experiments below $T_c$. [13–16] A pairing mechanism from these nonlocal and heavy fluxes is proposed, where it is shown that they can provide sufficient Bernoulli-like attraction force to compete with the Coulombic repulsion for the formation of Cooper pairs with a pair size of a few times of the Cu-Cu bond length in the $CuO_2$ plane.

**Nonadiabatic flux flows generated by anharmonically coupled phonons**

We first compute the charge densities with any two phonon modes anharmonically coupled together using DFT, following a previous method. [2] We then use our new method to determine the fluxes between ions to generate the flux pattern (See Methods). Our main computational focus is five representative parent undoped cuprate materials of $LaSrCuO_4$ (LSCO), $YBa_2Cu_3O_7$ (YBCO), $HgBa_2CuO_4$ (Hg-1201), $HgBa_2CaCu_2O_6$ (Hg-1212) and $HgBa_2Ca_2Cu_3O_8$ (Hg-1223), with the corresponding $T_{c,max}$ of 35 K, 92 K, 94 K, 128 K and 134 K, respectively, if at the optimal doping. We also compare the flux pattern of these cuprates with other related materials, by replacing Cu with Ni in LSCO, i.e., LaSrNiO$_4$ (LSNO), and with other types of superconductors, such as pressurized hydrides (H$_3$S), BCS (FeB$_4$, Nb$_3$Al, MgB$_2$) and Fe-based ones (FeSe, NdFeAsO).

**Figure 1a** shows a portion of Hg-1212 together with some solved fluxes, such as from Cu to apical O (O$_{ap}$), from Hg to O$_{ap}$, from Cu to O in the CuO$_2$ plane (O$_{in}$), and from O$_{in}$ to O$_{ap}$. We also notice strong fluxes to or from Ba, and between two nearest Cu ions inside the CuO$_2$ plane (Cu$_{in}$) or two nearest O$_{in}$ ions. Apart from fluxes associated with the bonds between Hg, Ba, Cu, Ca ions and their nearest neighboring oxygen ions, we also consider the flux between Ba ions, between Cu$_{in}$ ions along diagonal directions, and between oxygen ions along diagonal directions, when solving flux patterns. **Figure 1b** shows an example flux pattern for the net flux strength and direction in Hg-1212, flowing from Ba to O$_{ap}$ ions in the supercell for each coupled pair of phonons. **Figure 1c** shows that for any given mode from low to high frequencies, the average flux strength from Ba to O$_{ap}$ of all its coupled phonon modes, i.e., a projection of the flux pattern in Fig. 1b to 1-Dimension. We see a general trend in Hg-1201, Hg-1212 and Hg-1223 that the last few high frequency modes generate much larger flux strength than other modes, and an increasing trend with the number of CuO$_2$ layers.

Furthermore, we find that the sign of most fluxes in the flux pattern of Hg-1201, surprisingly, remains unchanged with reversed phonon oscillation phase in a period (i.e., from positive "+" phonon amplitude to the negative "-" one, **Supplementary Information Figure S1a**), when the phonon pairs are anharmonically coupled. Note that the coupling also leads to a lower energy than the uncoupled single mode cases. Therefore, it suggests that an unusual nonlocal net flux flow oscillation, rather than the trivial local flux oscillation, can be supported by anharmonically coupled phonons in cuprates. We thus further quantify this preference by calculating the flux pattern difference between positive and negative phonon amplitudes using the cosine similarity of flux patterns, i.e., smaller difference (higher similarity) suggests the nonlocal flux flow is preferred, while larger difference (lower similarity) indicates the local flux oscillation instead. Note that to analyze the generality of this effect, we also calculate the similarity between most types of ions in Hg-1201 (See Supplementary Information Fig. S2 for part of the original data). The results show that for most types of fluxes there is a strong tendency toward the flow. This tendency is also obviously more general if the fluxes are generated by anharmonically coupled phonon pairs than the uncoupled single mode cases of a higher energy (**Figure S1b**). To account for the fact that such fluxes are driven by the anharmonic phonon coupling, we define the "flux flow metric" as the average flux pattern similarity of all mode pairs, weighted by their corresponding phonon coupling strength (See Methods).

In **Figure 2a**, we compare the metric among a pressurized hydride ($H_3S$), five cuprates (Hg-1201, Hg-1212, Hg-1223, YBCO, LSCO), two iron-based superconductors (NdFeAsO, FeSe), three BCS superconductors ($MgB_2$, $Nb_3Al$, $FeB_4$), LSNO, and graphite. The materials are ranked with increasing superconducting transition temperature $T_c$ from bottom up, where $H_3S$ and all the high-$T_c$ cuprates, except for LSCO, show a strong tendendy toward nonlocal flux flow, while other materials show either the opposite tendendy (i.e., local oscillation) or no such a tendency. Note that as the only material under pressure in Fig. 2a, $H_3S$ calculated at 200 GPa shows the highest flux flow metric, while also having the highest $T_c$. In Fig. 2b, we further obtained the flux flow metric for Hg-1212 under different hydrostatic pressure. The increasing trend of flow metric correlated well with the experimentally observed increasing of $T_c$ under pressure. [17]

The above results thus likely suggest a unique picture of flux-phonon interactin in cuprates. In such an interaction the anharmonically coupled phonon serves as the flow generator, and simultaneously, the nonlocal flux flow oscillation serves as the phonon-coupling stabilizer. This is different from the conventional assumption that dynamic charge transfer is a simple collection of local flux oscillations, limited between nearest-neighbor ions and largely following the period of phonons. However, since phonons are the generator of these fluxes, the charge flux speed is regulated by phonon frequencies rather than following the Fermi velocity of electrons.

The nonlocal flux flow oscillation regulated by phonons thus introduces a new perspective for thinking about the pairing mechanism in cuprates, as upon doping extra charges the strength of such nonlocal flux can be further magnified well beyond the instrisic parent cases. We will demonstrate in the last section that the frequency difference between high and low frequency phonons, and hence the speed difference of fast and slow fluxes can provide a strong pairing strength through Bernoulli-like effect. Here we simply divide fluxes into fast and slow groups based on the comparison with the average phonon frequency, and define the so-called flux pairing strength $G$ as the difference between the two groups for the weighted average of squared frequency $G \stackrel{\text{def}}{=} \tilde{f}_{fast}^2 - \tilde{f}_{slow}^2$ (See Methods). $G$ for five cuprate families are shown in **Figure 2c**, where the meterials are ranked with increasing $T_{c,max}$. We find that $G$ (blue bars) also monotonically increases. We further decompose the flux pairing strength $G$ by the contribution of each type of flux (Supplementary Information **Figure S3**), and find that a large portion of the $T_{c,max}$ dependence is from Ba (or Sr) related fluxes. The flux strength to Ba (Sr) ion of last four high frequency phonon modes are thus also presented as orange bars in **Figure 2c**, which, together with the $G$ bar, shows a good correlation with $T_{c,\text{max}}$.

**Dynamic flat band for heavy fluxes**

The fluxes analyzed in the real space above should correspond to the dynamic energy band in the reciprocal space. That is the energy band change perturbed by phonons. Previously, we found that such a band change in DFT simulations are very responsive at the $k$ point (0.5, 0.5, 0) for cuprates [2]. Thus here we focus on the free-hole-like parabola at the top of this $k$ point in DFT to estimate the energy scale of the dynamic band, as illustrated in **Figure 3a**. The original band $E_k = ak^2 + b$ is changed to $E'_k = a'k^2 + b'$ by phonons, so that for $\Delta E_k = \Delta a k^2 + \Delta b$, we have $\Delta a = a' - a$ and $\Delta b = b' - b$. The $\Delta a$ induced by each phonon mode in cuprates shows that the largest change is mostly in the high frequency range (SI **Figure S4**).

**Figure 3b** shows the $\Delta a$ for the three Hg-families by taking the average of the last four high-frequency modes that mainly contribute to the bands around Fermi level. With $\Delta a$ ranging from 0.05 to 0.12, the dynamic band is an order of magnitude flatter than the static energy band, as also illustrated in Fig. 3a. From Fig. 3b, $\Delta a$ decreases from Hg-1201, Hg-1212 to Hg-1223, suggesting that the dynamic band also becomes flatter with increasing number of $CuO_2$ layers. Importantly, the effective mass of charge fluxes in the dynamic flat band, $m_f = \hbar^2/(d^2\Delta E/dk^2) = \hbar^2/2\Delta a = 3.82/\Delta a\, m_e$, is thus in the range of $30 \sim 80 m_e$, giving a heavy flux with the speed limit set by the high frequency phonons. Also note that since for cuprates the oxygen anion framework sets the high frequency at around 70 meV, the heavy and "fast" charge flux is actually ten times slower than the Fermi velocity of electrons in cuprates.

In our previous work, [2] we proposed a two-band model to describe the behavior of charge carriers in cuprates in the underdoped region, with a spatially more delocalized shell component and a spatially more localized core component, from which the ARPES simulations agree well with experimental ARPES above $T_c$. We show here that the interaction term $U_f$ between the two components can be proved to be proportional to $cosk_x - cosk_y$ (Supplementary Information), which decides the pseudogap and Fermi arc in the ARPES simulation (See Methods for more details).

It has been observed universally for cuprates that below $T_c$ an additional flat band emerges just below the Fermi level in many ARPES experiments, [13–16] suggesting that the band is likely involved in the pairing process. Here we assume that when two holes approach, the interaction will force some portion of the hole charge to move from the shell component to the dynamic flat band, behaving as the heavy fluxes. We thus propose a three-band Hamiltonian by incorporating this flat band (See Methods). The ARPES simulation shown in **Fig. 3c** based on the three-band model clearly exhibits a flat band with high intensity around $k_x = \pm \frac{0.25}{\pi}$, in a good agreement with ARPES measurements below $T_c$. [13–16] **Figure S5** shows the simulated band structure below the Fermi level, where the flat band (blue) has a significant contribution from the heavy flux state (dot-dashed curve).

**Pairing by heavy-flux flow induced Bernoulli effect**

We now propose a pairing mechanism based on the heavy flux flow. When two holes in cuprates approach each other (upper part of **Fig. 4a)**, the electric repulsion between the two positive cores creates a narrow valley in the electric potential landscape between the two. Faster fluxes will prefer to flow through this narrow region, while slower fluxes will prefer to flow in the outside broad region. The speed difference between the inner fast flux with a speed of $v_{fast}$ and the outside slow flux with a speed of $v_{slow}$ thus in principle can provide a Bernoulli attraction force $F_B$ to balance the Coulombic repulsion force $F_C$ (lower part of Fig. 4a), paring the two holes into a Cooper pair.

We further build a model to estimate the magnitude of the paring strength (See Methods). It can be shown that the Bernoulli force $F_B \propto 2m(v_{fast}^2 - v_{slow}^2)/l$, as illustrated in Fig. 4a, where $v_{fast}$ and $v_{slow}$ can be estimated from DFT (e.g. Fig. 1c) and $l$ is the pair separation. Another important balancing factor between $F_B$ and $F_C$ is that when a certain portion $x$ of the hole charge contributes to the heavy flux, only $(1-x)$ of each hole's charge will be responsible for the Coulomb repulsion. **Figure 4b** shows the energies $E_B$ and $E_C$ as well as the total energy at $x = 0.3$ versus distance $l$, which shows a local energy minimum of around -70 meV at a pair equilibrium distance $l_{eq}$ around 15 Å. When two holes interact, **Figure 4c** shows that the solved balancing flux portion $x$ increases with decreasing $l_{eq}$ of the pair. For a wide range of flux dynamic effective mass $m_f$ (estimated in the range of $30 \sim 80 m_e$ in Fig. 3), it requires less than 30% ~ 40% flux portion from a hole in order to show a balancing distance of around 15 Å.

We can now see that the flux pairing strength $G$ we defined earlier uses the key component of the Bernoulli force, $v_{fast}^2 - v_{slow}^2$, which is proportional to the difference of squared frequency, $f_{fast}^2 - f_{slow}^2$. This pairing mechanism is thus supported by the material dependence of $T_{c,max}$ with the flux paring strength presented in **Figure 2c.** Interestingly, Figure 4c also suggests that the flux portion $x$ may increase significantly toward overdoping, as the hole-hole separation may be forced to be smaller than $l_{eq}$ beyond a certain doping concentration. This implies a potential connection with the universal Planckian dissipation and $T$-linear resistivity that is prominent in overdoped cuprates, [18–20] through certain scattering of flux flows based on this picture. As far as we know, this is the first time that such a flow picture related to the macroscopic Bernoulli effect is applied to the microscopic pairing mechanism in cuprates with a clear material dependence. Our work thus could also form the foundation of the high-throughput search and machine learning design of new superconductors with breakthrough $T_c$ at ambient pressure in the future.

**Contribution**

X.L designed and supervised the research. X. L conceived the physical picture of heavy flux flow and proposed the pairing mechanism. X.C performed the DFT simulation and the modelling for pairing mechanism. J.P and X.C performed the flux analysis. J.D and C.L constructed the 3-band model and J.D performed the ARPES simulation. Y.C participated in the DFT calculation of FeSe and NdFeAsO. All authors discussed the results and wrote the manuscript.

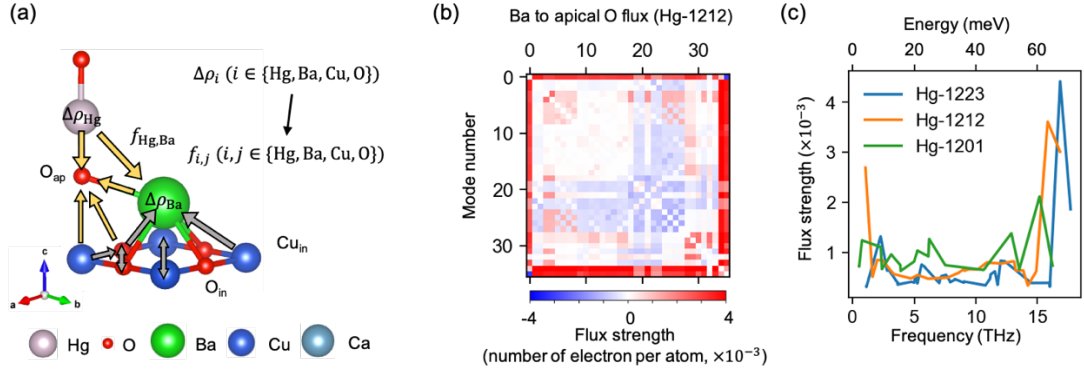

**Figure 1**. **Charge fluxes generated by anharmonic phonons in DFT simulations.** **(a)** Illustration of the flux model, where charge difference $\Delta\rho$ of each atom obtained using Bader analysis is assigned to each node, and nonzero fluxes are assigned to nearest neighboring atoms and several other farther pairs. Fluxes are illustrated as arrows. The entire set of nonzero fluxes considered is in Supplementary Information. The arrow directions correspond to the preferred flux directions for high and low frequency modes, which are the same for Hg-1201, Hg-1212 and Hg-1223. **(b)** Flux strength from Ba to apical oxygen for parent undoped Hg-1212. Each square represents the flux strength induced by the corresponding pair of phonon modes coupled anharmonically. Red (positive value) means the flux flows from Ba to apical oxygen, and blue (negative) value means the opposite. **(c)** Mean value of Ba to apical oxygen flux strength for each phonon mode averaged from all its anharmonically coupled modes for Hg-1201, Hg-1212 and Hg-1223. The value is calculated by summing over the matrix values in (b) along one dimension.

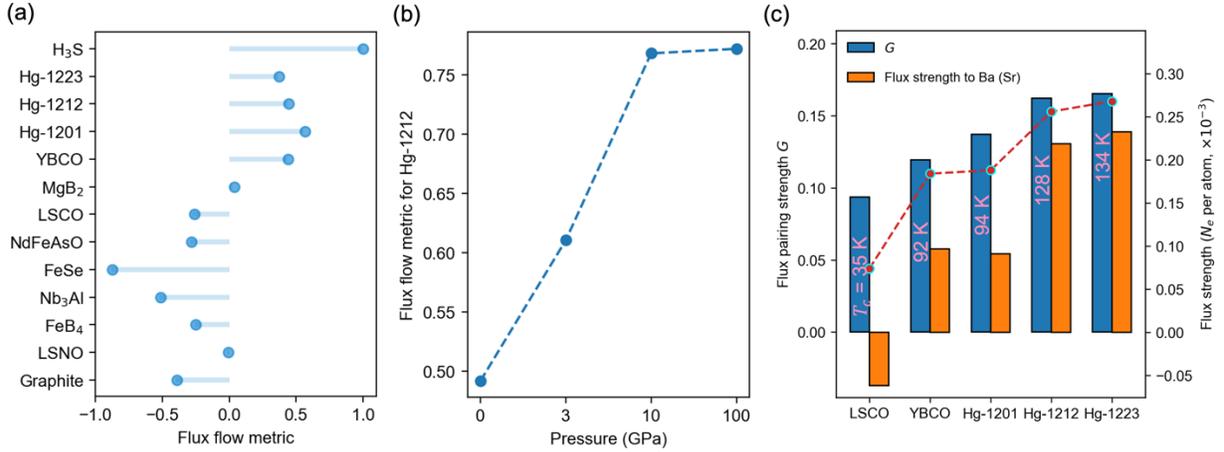

**Figure 2. Unique flux flow behavior in cuprate superconductors. (a)** Flux flow metric for various materials, defined as the similarity of flux patterns for main types of atom connections between "+" and "-" amplitudes in an oscillation period for coupled phonon pairs, weighted by the phonon coupling strength. The calculation of $H_3S$ is under 200 GPa pressure. LSNO corresponds to replacing Cu with Ni in LSCO. **(b)** The flux flow metric for Hg-1212 under pressure. **(c)** Flux pairing strength $G$ and the flux strength to Ba or Sr for LSCO, YBCO, Hg-1201, Hg-1212, and Hg-1223. Red dots are the corresponding $T_{c,max}$ for these materials. For a better visualization, the value of the flux strength to Ba or Sr is the average flux strength of the last four modes (absolute value) minus the average value of all modes.

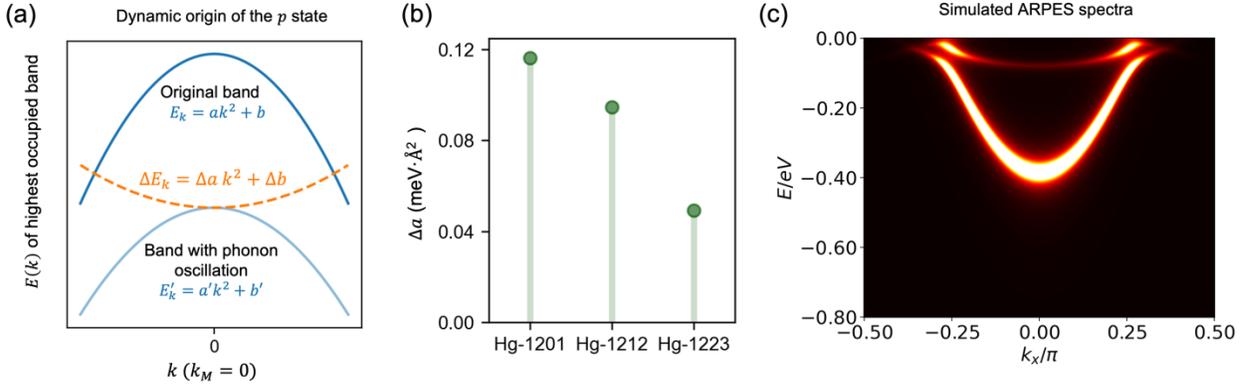

**Figure 3. ARPES simulation using the 3-band model. (a)** Illustration of the origin of the heavy flux state from the dynamic change of the highest occupied band around the *k*-point (0.5, 0.5, 0) perturbed by phonons. **(b)** $\Delta a$ of the three Hg-family materials from an average over the last four phonon modes from DFT simulations. **(c)** Simulated ARPES spectra with $k_y = 0.85\pi$.

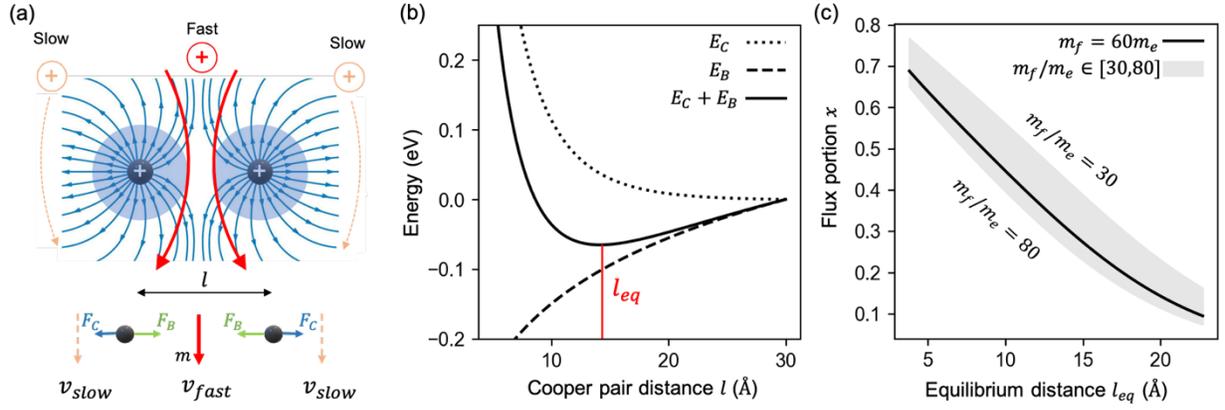

**Figure 4. A pairing mechanism from the Bernoulli force generated by heavy flux flow. (a)** Schematic illustration of the electric field lines (blue curves) generated by two approaching holes. Red and orange arrows indicate that some fluxes tend to flow faster between the two holes and slower in the outside region. The bottom part illustrates the Bernoulli force $F_B$ balanced by the Coulomb force $F_c$. **(b)** Total energy of the system as a function of hole-hole distance $l$ from the Bernoulli energy $E_B$ and the electrostatic repulsion energy $E_c$ with flux portion $x = 0.3$ and flux dynamic effective mass $m_f = 60 m_e$. **(c)** Equilibrium hole-hole distance $l_{eq}$ corresponding to different flux portion $x$ and flux mass $m_f/m_e \in [30, 80]$.